\documentclass[11pt, a4paper]{article}

\usepackage[dvips]{graphicx}
\usepackage{listings}
\usepackage{hyperref}

\def\BibTeX{{\rm B\kern-.05em{\sc i\kern-.025em b}\kern-.08em
    T\kern-.1667em\lower.7ex\hbox{E}\kern-.125emX}}

\def\MARU#1{\leavevmode \setbox0\hbox{$\bigcirc$}%
\copy0\kern-\wd0 \hbox to\wd0{\hfil{#1}\hfil}}

\begin{document}
\title{Privacy-Preserving Infection Exposure Notification
without Trust in Third Parties}
\author{Kenji Saito, Mitsuru Iwamura
\thanks{The authors are with Graduate School of Business and Finance,
Waseda University, email: ks91@aoni.waseda.jp}
}
\date{}

\maketitle

\begin{abstract}
In response to the COVID-19 pandemic, Bluetooth-based contact tracing has been
deployed in many countries with the help of the developers of smartphone
operating systems that provide APIs for privacy-preserving exposure
notification.
However, it has been assumed by the design that the OS developers, smartphone
vendors, or governments will not violate people's privacy.

We propose a privacy-preserving exposure notification under situations where
none of the middle entities can be trusted.
We believe that it can be achieved with small changes to the existing
mechanism: random numbers are generated on the application side instead of the
OS, and the positive test results are reported to a public ledger (e.g.
blockchain) rather than to a government server, with endorsements from the
medical institutes with blind signatures.
We also discuss how to incentivize the peer-to-peer maintenance of the public
ledger if it should be newly built.

We show that the level of verifiability is much higher with our proposed
design if a consumer group were to verify the privacy protections of the
deployed systems.

We believe that this will allow for safer contact tracing, and contribute to
healthier lifestyles for citizens who may want to or have to go out under
pandemic situations.

\begin{description}
\item[Keywords:]
Contact tracing, Exposure notification, Privacy,\\
Blind signature, Blockchain
\end{description}

\end{abstract}

\section{Introduction}
\subsection{Motivation}
Contact tracing is a technique traditionally used by public health authorities
to combat infectious diseases, which until now has relied primarily on manual
methods.
It is based on the concept of ascertaining others with whom the infected
person has come into contact while there is a possibility of spreading the
infection to others.
The possibility of close contact with an infected person is notified to the
person with whom the contact has been made, so that appropriate safety
measures can be taken, such as self-quarantine and/or testing.

During the ongoing pandemic of COVID-19, consideration was given to applying
digital communication technology to support and massively scale these efforts.
By embedding proximity-detection functionality into mobile devices, it is
considered possible to identify past close contacts of people who later test
positive, and send them notifications with instructions on next steps.
Health authorities can use this information to control the spread of the
disease.

With this in mind, Bluetooth-based exposure notification (contact tracing) has
been proposed.
Applications have been developed in many countries for use on smartphones with
the help of the developers of smartphone operating systems, namely Google and
Apple, who provide APIs for privacy-presereving exposure notification.
However, in the designs of these applications, it has been assumed that the OS
developers, smartphone vendors, or governments will not violate people's
privatey.
We believe that better privacy protection without relying on the integrity of
these trusted third parties is needed.

\subsection{Contributions}
Contributions of this work are as follows:
\begin{enumerate}
\item We identified threats that intermediaries could pose, alone or in
collusion, with respect to privacy protection of the users in the existing
design of the OS-assisted exposure notification.

\item We proposed a design to mitigate the threats by minimal changes to the
existing design, along with an incentive design for the operation of the
system in case a verifiable public ledger should be newly built independently
of the government authority.

\item We showed that the level of verifiability is much higher with our
proposed design if a consumer group were to verify the privacy protections of
the deployed systems.
\end{enumerate}

\subsection{Organization of This Article}
The rest of this articile is organized as follows:
section \ref{sec-background} gives brief background information to better
understand our proposal: exposure notification design by Google and
Apple, Merkle accumulator, blind signature, and blockchain.
Those readers who are familiar with the mentioned technology can move directly
to section \ref{sec-problem}, which gives the problem statement and enumerates
threats.
Section \ref{sec-design} describes our proposal to change the existing design.
Section \ref{sec-evaluation} compares our proposed design with the original one
with respect to verifiability of privacy protection.
Section \ref{sec-related} explains some related work.
Finally,
section \ref{sec-conclusions} gives conclusive remarks.

\section{Background}\label{sec-background}
\subsection{Exposure Notification Design by Google and Apple}
\begin{figure*}[h]
\begin{center}
\includegraphics[scale=0.48]{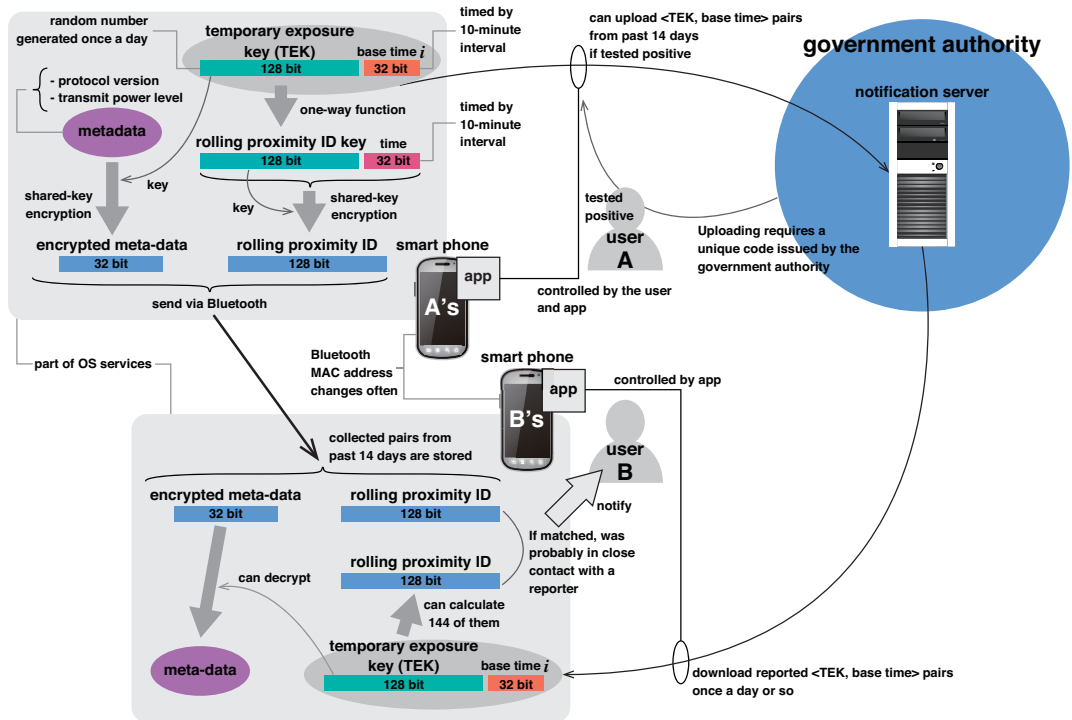}\\
{\small
\begin{itemize}
\item[*] The application is developed by the government.
\end{itemize}
}
\caption{Overview of OS-assisted exposure notification.}\label{fig-gaen}
\end{center}
\end{figure*}

Google and Apple provide exposure notification as part of their OS
services\cite{AppleGoogle2020:ENFAQ},
as roughly illustrated in Figure \ref{fig-gaen}.
This design is often referred to as GAEN (Google Apple Exposure Notification)
in literatures.
We follow the convention hereafter.
The cryptographic specification\cite{AppleGoogle2020:ENCrypto},
Bluetooth specification\cite{AppleGoogle2020:ENBluetooth} and
the programming framework\cite{Google2020:ENAPI}\footnote{This reference is
for Google Android API.} of GAEN have been published.

In GAEN, the time is numbered in 10-minute intervals, starting at 00:00:00 UTC
on January 1, 1970 (Unix epoch).

The device generates one random {\em temporary exposure key (TEK)} (128 bits)
every 24 hours.
It is valid for 144 time units (24 hours) starting from time $i$.
The device stores up to 14 TEKs (for two weeks) with their respective starting
time $i$.

The device generates the {\em rolling proximity ID key} (128 bits) from the
TEK of the day using a one-way function.
Along with the timing when the MAC address of Bluetooth LE is changed,
a {\em rolling proximity ID} (128bit) is generated by encryption from the 
rolling proximity ID key and the time, which is sent by Bluetooth
communication as a beacon.
Other smartphones listen to these beacons, storing them upon receiving them,
and broadcast their own beacons as well.

The system can also attach encrypted metadata to the beacon, such as protocol
version and transmit power level, which can be decrypted with (the key
generated from) the TEK of the day.

The device owned by a person who tests positive and reports voluntarily sends,
for example, the TEKs for the past 14 days and their starting time $i$ to the
notification server.

All users will periodically download those reported data (e.g. once a day).
The user's device can recalculate the rolling proximity ID from the TEK and
$i$.
One TEK can create 144 rolling proximity IDs.
If the same rolling proximity ID obtained by the calculation is stored on your
device, then you have a high probability that you were in close contact.

If the device has received the encrypted metadata along with the rolling
proximity ID, it can decrypt the data using the TEK, although there is no
guarantee that the information is correct.
This means that we must also consider the existence of other applications
and/or malware that perform the same Bluetooth communication.

In the designs of these applications, it has been assumed that the OS
developers, smartphone vendors, or governments will not violate people's
privatey.

\subsection{Merkle Accumulator}

\begin{figure}[h]
\begin{center}
\includegraphics[scale=0.4]{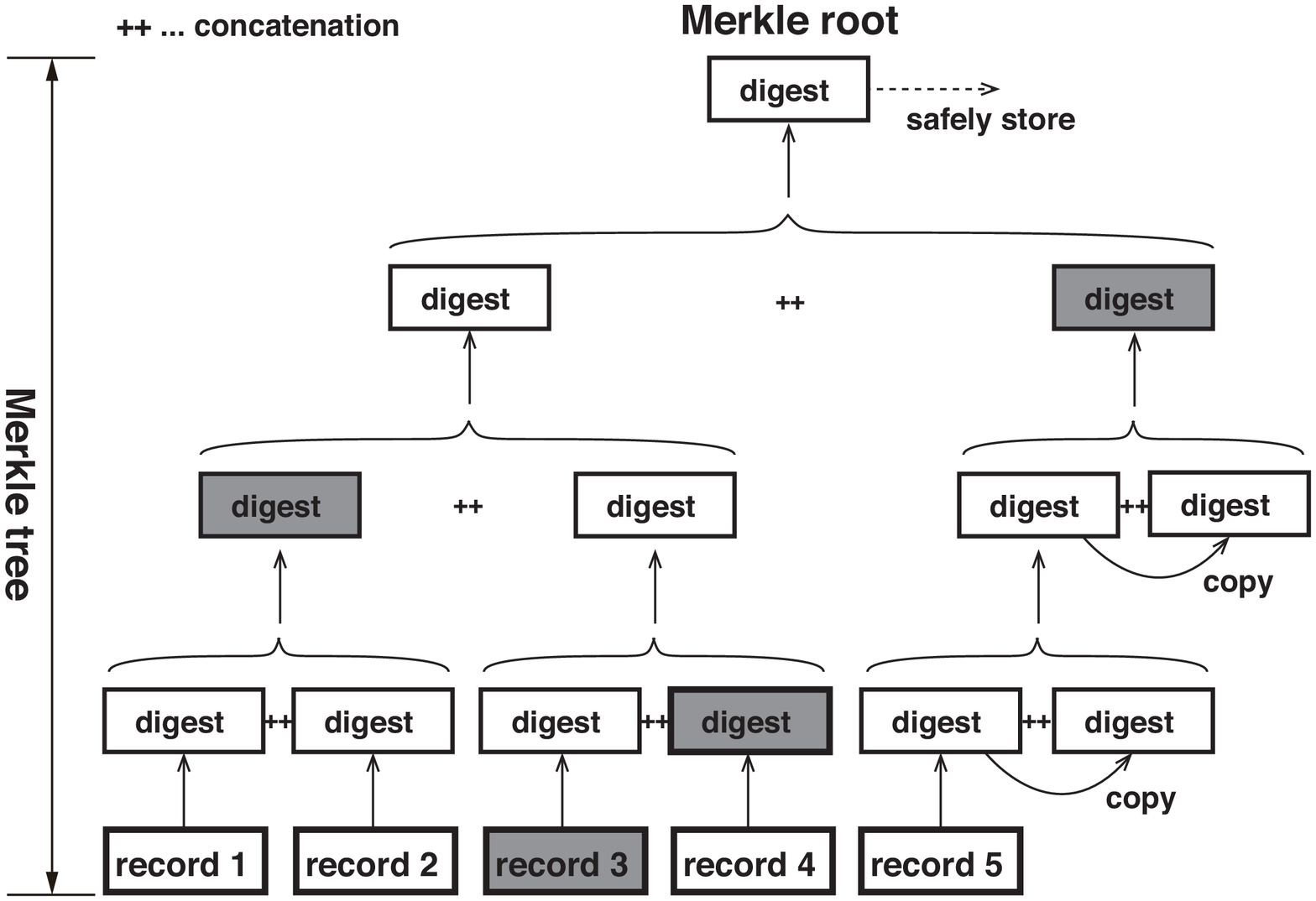}\\
{\small
\begin{itemize}
\item[*]
In order to confirm the existence of record 3, see if the same Merkle
root as safely stored can be calculated from the provided partial tree
(Merkle proof) shown in gray.
\end{itemize}
}
\caption{Merkle tree and Merkle proof.}\label{fig-merkle-tree}
\end{center}
\end{figure}

A Merkle tree\cite{Merkle1988:Tree} is a hash tree structure based on a
cryptographic hash function that produces cryptographic digests.
Such a tree allows representation of multiple elements with a single value,
and is used for proof of existence of elements while obscuring others, as
illustrated in Figure \ref{fig-merkle-tree}.

A Merkle accumulator is a Merkle tree to which elements can be added
incrementally, to accumulate evidences of records.

In our work, we use a Merkle accumulator to record with verifiable evidences
the reports from people tested positive.

\subsection{Blind Signature}

\begin{figure}[h]
\begin{center}
\includegraphics[scale=0.36]{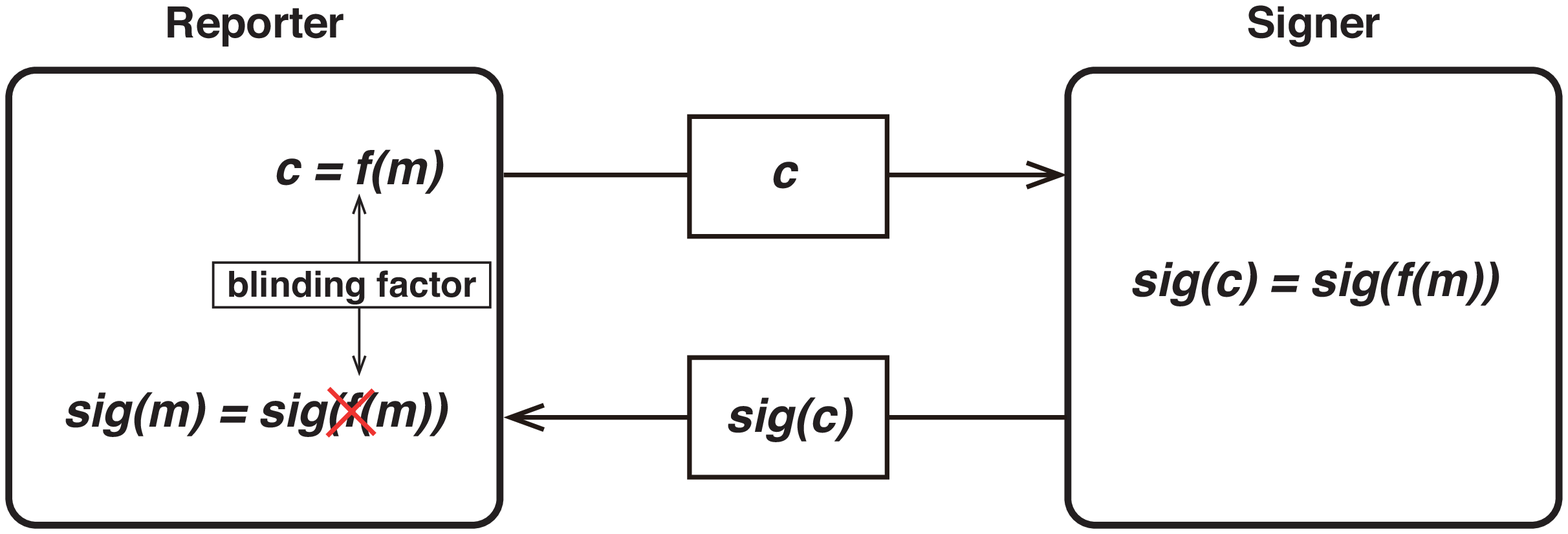}\\
{\small
\begin{itemize}
\item[*] The reporter wants to have message $m$ signed by the signer
without revealing $m$.
They first wrap $m$ with blinding factor $f$, and send it to the signer for
signing.
After receiving the signature, they remove $f$ from it to obtain the signature
for $m$ verifiable with the signer's public key.
\end{itemize}
}
\caption{Blind signature.}\label{fig-blind-signature}
\end{center}
\end{figure}

Blind signature\cite{Chaum1983:BlindSignature}
is a technique for digitally signing hidden data as if it were signed
blindfolded, as illustrated in Figure \ref{fig-blind-signature}.
It was developed to enable anonymous electronic payments.

In our work, we apply blind signatures to reports from people tested positive,
signed by the medical institute that performed the test.

\subsection{Blockchain}
Blockchain is a structure introduced for realization of
Bitcoin\cite{Nakamoto2008:Bitcoin}, a digital cash system.
It allows for tamper-evident storage verifiable by the public so that it can
provide verifiability of digital signatures in the
past\cite{Saito2020:Blockchain} that is otherwise difficult because of
possible compromise of private keys or signature algorithms, or expiration of
public key certificates.

\begin{figure}[h]
\begin{center}
\includegraphics[scale=0.37]{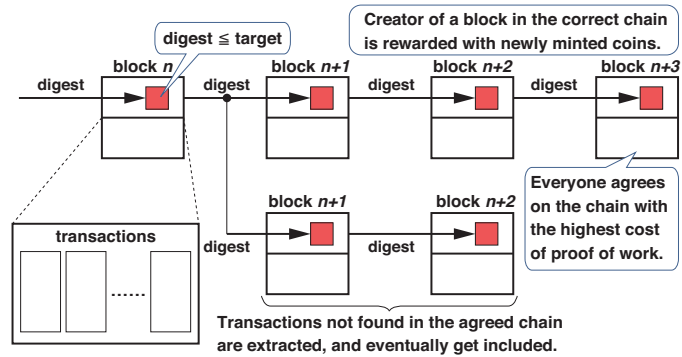}\\
\caption{Blockchain based on proof of work.}\label{fig-blockchain}
\end{center}
\end{figure}

As Figure \ref{fig-blockchain} shows, each block contains the cryptographic
digest of the previous block.
Such a digest must meet a certain criterion; it needs to be less than or equal
to the pre-adjusted and agreed target stored in or calculated from the block.
Since the digest is calculated by a one-way function whose outputs are evenly
distributed, no one can intentionally configure a block to satisfy the
criterion.
Instead, they need to partake repetitive trials to change the values of some
nonce in the block they are creating until they get a right digest.

The necessity of repetitive trials functions as a {\em proof-of-work}
mechanism intended to be a protection against falsification.
A transaction itself cannot be falsified unless digital signatures are
compromised.
But it is conceivable to remove some transactions from a past block or to add
fabricated transactions that did not exist at the time to it.
If one tries so, the digest of the block is changed and is typically greater
than the target.
Then they would have to retry the proof of work for the block.
This changes the digest stored in the next block, which in turn means that the
digest of the next block is also changed and is typically greater than the
target, and so on.
In short, ones with a malicious intention would have to redo the proof of
work from where they want to change, and outdo the ongoing process of adding
blocks eventually to make the change valid, which has generally been considered
highly difficult.

Such proof of work limits the number of proposed blocks at one time.
But there still is a possibility of multiple participants each proposing a
new block at roughly the same time, which may be accepted by different sets of
participants.
Then the chain may have multiple ends that are extended independently
from one another, resulting in a fork of the blockchain with multiple (and
possibly contradicting) histories of blocks.
If this happens, the branch that is the most difficult to produce (or rewrite)
is chosen by all participants, which is the branch with the most accumulated
proof of work.
This mechanism, called {\em Nakamoto consensus}, tries to enforce
that the most difficult chain branch to falsify is chosen as the single
correct history\footnote{For imperfection of the design of Nakamoto consensus,
readers are referred to a past work\cite{Saito2016:Blockchain} by the first
author of this paper.}.

Ethereum\cite{Buterin2013:Ethereum} is a blockchain-based application platform
to assure authenticity of program codes ({\em smart contracts}), their
execution logs and the resulted states.

Both Bitcoin and Ethereum are based on proof of work, which is protected by
the high power costs associated with it, but these costs are known to balance
in the long run with the market value of the respective native currencies
earned through block creation\cite{Iwamura2019:BitcoinMonetaryPolicy}.
In other words, these blockchains are protected by the high market value of
their respective cryptocurrencies.

In order to avoid the environmentally burdensome power costs of proof of work,
recent blockchains such as Ethereum 2.0\cite{Ethereum:eth20Spec} and
Polkadot\cite{Wood2016:Polkadot} have adopted the idea of {\em proof of stake},
in which legitimate history is determined by weighted voting with a deposit of
native currency.
However, then again, the high market values of the respective native
currencies are still the major factors that keep these blockchains safe.

Both Ethereum 2.0 and Polkadot allow multiple private ledger applications to
be {\em anchored} to their central blockchains in the form of {\em shards} and
{\em parachains}, respectively.

In our work, we store Merkle roots of reports in blockchain.

\section{Problem}\label{sec-problem}

This work proposes a privacy-preserving exposure notification mechanism
that tolerates situations where none of the middle entities can be trusted not
to violate people's privacy.
The solution must mitigate the following threats to privacy of the users that
are present in GAEN, where {\em private data} denotes that of the phone user,
such as the phone number, e-mail address, physical location, real name, etc.

\begin{enumerate}
\item OS developer and/or smartphone vendor alone can :
\begin{enumerate}
\item encode private data or a marker in a TEK.
\label{pro-tek}
\item send an arbitrary beacon containing private data or a marker.
\label{pro-beacon}
\item encrypt private data or a marker as the associated metadata.
\label{pro-metadata}
\item collect identities of the close contacts with the user associated with
such private data or a marker.
\label{pro-collect-beacons}
\item notify exposures falsely to any specific users to stop or slow down
their social activities.
\label{pro-exposures}
\end{enumerate}

\item The government alone can :
\begin{enumerate}
\item collect identities of the reporters.
\label{pro-collect-reporters}
\item stop or slow down social activities of political enemies, for
example, by bringing agents close to them, and having the agents later report
falsely. 
\label{pro-agents}
\end{enumerate}

\item OS developer and/or smartphone vendor and the government can collude to :
\begin{enumerate}
\item make fake reports from specific users, and have them downloaded by
general public, to stop or slow down social activities of the close contacts
with the users.
\label{pro-disable}
\end{enumerate}

\end{enumerate}

Threats \ref{pro-tek}, \ref{pro-beacon} and \ref{pro-metadata} are possible
because generation of TEKs, deriving rolling proximity IDs from them, and
metadata are processed within the OS services, and the OS can send arbitrary
beacons anyway, which is not detectable as applications do not know the values
of TEKs until the users decide to report.
Threats \ref{pro-collect-beacons} and \ref{pro-exposures} are possible because
what is performed as a result of receiving a beacon is hidden within the OS
services.

Threat \ref{pro-collect-reporters} is possible because it is the government
authority that issues the unique code to the person tested positive, and it
can be designed to map the code to the person.
Threat \ref{pro-agents} is possible because it is the government authority
that processes the reports, and they can allow agents to bypass the normal
reporting procedures to send their TEKs to the server, as the application is
developed by the government, possibly with some hidden features.

Moreover, threat \ref{pro-disable} is possible because the government
authority can obtain the TEKs of a specific person from OS developers or
phone vendors\footnote{According to Google API, applications cannot obtain
TEKs without displaying a dialog that requests consent from the user.
Therefore the government needs cooperations from the OS developer or phone
vendor if they want to obtain the TEKs without letting the user know it.}, and
store them on the server.

\section{Design}\label{sec-design}
\subsection{Basic Design}

We believe that the above threats can be mitigated with the following three
small changes to the existing mechanism, as illustrated in
Figure \ref{fig-en-wo-ttp}:

\begin{figure*}[h]
\begin{center}
\includegraphics[scale=0.48]{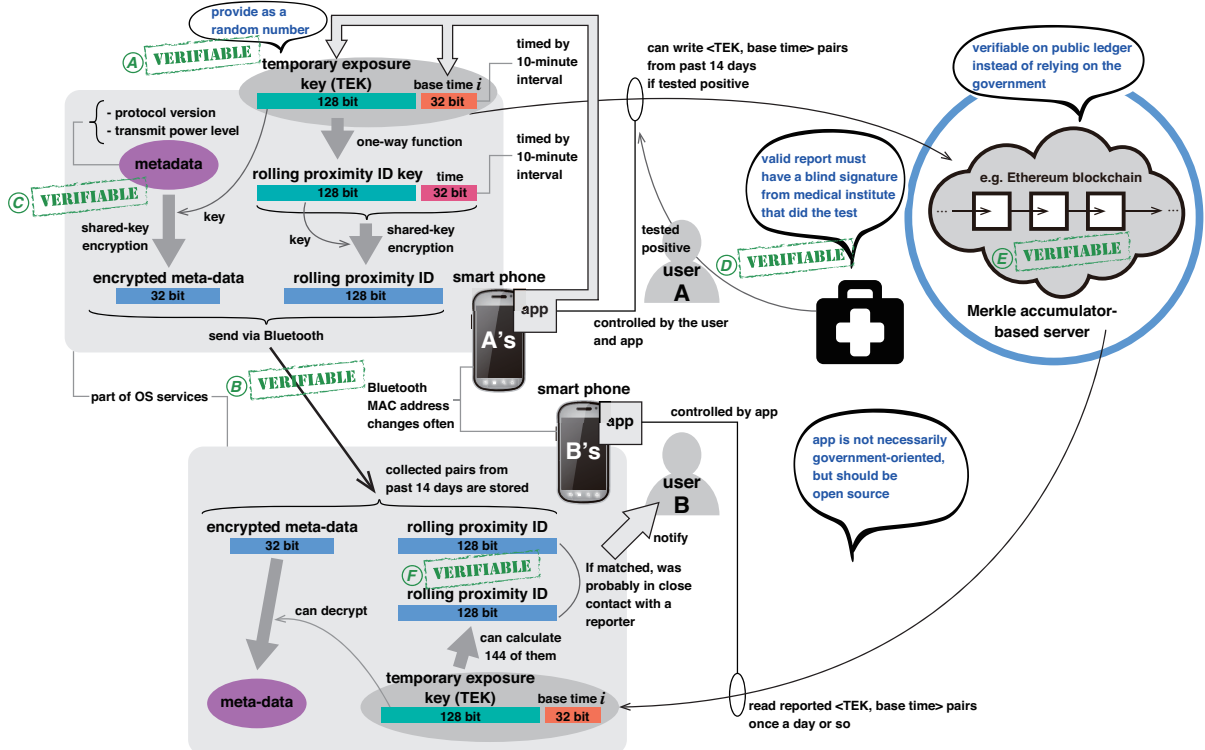}\\
\caption{Overview of exposure notification without trust in third parties.}\label{fig-en-wo-ttp}
\end{center}
\end{figure*}

Furthermore, the applications should be open source so that they can be built
and installed on smartphones by the phone users if they want.

\subsubsection{Generation of TEKs at the Application Side}
We change the distribution of the functionality so that TEKs are generated as
random numbers on the application side instead of within the OS services.

\subsubsection{Endorsement by Medical Institutes with Blind Signatures}
The positive test results must be reported with endorsements (blind
signatures) from the medical institutes who did the test.
In order to avoid correlation, each $\langle$TEK, base time$\rangle$ pair is
blind-signed independently.
The uploaded report takes the form of $\langle$TEK, base time, signature,
MIID$\rangle$ for each TEK, where MIID is the medical institute ID.
The public key certificate used for verifying the signature can be retrieved
with MIID as the search key from a public database.

Typically, when a person is informed of a positive test result at a testing
facility, they are also informed of a short-term URL via a QR code, for
example.
When the person accesses the URL on their smartphone, the exposure notification
application sends 14 blindfolded $\langle$TEK, basetime$\rangle$ pairs to the
medical institute's server, which returns blind-signed results to the
application.
The person can then choose to report them.

\subsubsection{Storage of Reports in a Verifiable Public Ledger}
The positive test results are reported to a verifiable public ledger
(e.g. blockchain) rather than to a government server.

Assuming blockchain is used as the core of a verifiable public ledger, Merkle
accumulator is applied by the server to lower the cost (transaction fees and
processing time) of using blockchain.
To prevent correlation, reports arriving at close timing are mixed with other
reports and shuffled before being added to the accumulator.
A Merkle tree is finalized every 24 hours, for example, to form a set of
reports to be downloaded by the application.
Our method is roughly the same as that presented in \cite{watanabe2020proof},
using a similar or the same smart contract for Ethereum as shown in
Figure ~\ref{fig-anchor}, which returns the number of the block when the Merkle
root was written, so that we can know approximately what time the root was
stored.

\begin{figure}[h]
\begin{center}
{\tiny
\begin{lstlisting}[frame=single]
contract Anchor {
  mapping (uint256 => uint) public _digests;
  constructor () public {
  }
  function getStored(uint256 digest) public view returns (uint block_no) {
    return (_digests[digest]);
  }
  function isStored(uint256 digest) public view returns (bool isStored) {
    return (_digests[digest] > 0);
  }
  function store(uint256 digest) public returns (bool isAlreadyStored) {
    bool isRes = _digests[digest] > 0;
    if (!isRes) {
      _digests[digest] = block.number;
    }
    return (isRes);
  }
}
\end{lstlisting}
}
{\footnotesize
\begin{itemize}
\item[*] store() saves the current block number for a stored digest.
\item[*] For a given digest, getStored() returns the block number if it is
stored.
It returns 0 otherwise.
\end{itemize}
}
\caption{Sample Anchoring smart contract code.}\label{fig-anchor}
\end{center}
\end{figure}

Normally, an application can download the set of reports that have been added
since the last time, and use all of them to create a single Merkle tree to
verify that the Merkle root is stored in Ethereum (at a plausible time).
If the application wants to check the existence of individual reports
separately, the server needs to provide a Merkle proof (partial tree), but
perhaps that would not be the case.

\subsection{Incentive Design}
We also discuss how to incentivize the peer-to-peer maintenance of the
public ledger, in case it is newly built and specifically used for the
purpose of exposure notification.
However, a private ledger would not provide sufficient proof due to a
relatively large margin for malicious involvement.
Therefore, it would be more appropriate to provide the new ledger in the form
of an Ethereum 2.0 shard or Polkadot parachain, for example, which can be
anchored firmly to existing blockchain.

Blockchain is originally designed to collect transactions, form a Merkle
tree out of them, and store them in a verifiable block.
By replacing transactions with reports, we obtain a straightforward design for
the new ledger.

We would probably like to keep the structure of a traditional blockchain,
where there is a reward for the creation of a block to incentivize its
maintenance, but preferably without a fee for writing a report transaction.
We also want to keep the market price of the rewards stable, given that we are
protected by the high market price of the currency, although we do not need to
maintain it for a long time given the expected duration of the pandemic.
The authors of this work have proposed a way to stabilize the price of
such a cryptocurrency and at the same time to abandon transaction fees
\cite{Saito2019:Stable},
which may be applicable to the design.

\section{Evaluation}\label{sec-evaluation}
\subsection{Verifiability and Privacy}
We evaluate our proposal by a thought experiment: if a consumer group were to
verify the privacy protections of the two different systems, original GAEN and
our proposed versions, the results would be as shown in Table~\ref{tab-eval}.

\begin{table*}
\begin{center}
\caption{Evaluation if a consumer group were to test the applications.}
\label{tab-eval}
{\small
\begin{tabular}{l|l|l|l}\hline
Point of weakness&
Corresponding&
Our proposal&
GAEN
\\
&
threats&
&
\\\hline

{\em \MARU{A} Generation of TEKs}&
\ref{pro-tek}, \ref{pro-collect-beacons}&
Prevented&
Undetectable\\\hline

{\em \MARU{B} Content of beacons}&
\ref{pro-beacon}/\ref{pro-collect-beacons}&
Detectable/Suspectable&
Undetectable\\\hline

{\em \MARU{C} Metadata}&
\ref{pro-metadata}/\ref{pro-collect-beacons}&
Detectable/Suspectable&
Undetectable\\\hline

{\em \MARU{D} Reporting}&
\ref{pro-collect-reporters}, \ref{pro-agents}, \ref{pro-disable}&
Made more difficult&
Undetectable\\\hline

{\em \MARU{E} Stored reports}&
\ref{pro-disable}&
Detectable&
Undetectable\\\hline

{\em \MARU{F} Matching proximity IDs}&
\ref{pro-exposures}&
Detectable&
Detectable\\\hline

\end{tabular}
}
\end{center}
\end{table*}

For testing our proposed version, first, the consumer group modifies the
open-source application to do the necessary logging, install it on their
smartphones, and then perform testing.
Below, we explain how our proposal works for each point of weakness.

\begin{description}
\item[\MARU{A} Generation of TEKs:]
In our proposal, the application generates the TEKs, which prevents the OS
from encoding any information into them.
Naturally, the beacons cannot carry any information resulting from encoding
in TEKs either.

\item[\MARU{B} Content of beacons:]
The rolling proxmity ID can be reproduced if the TEK and base time are known,
to verify that the contents sent by Bluetooth are correct.
If a value that is not based on the TEK is sent, it can be detected, and if
it is detected, it can be suspected that a secret process is embedded in the
receiver's OS.

\item[\MARU{C} Metadata:]
Encrypted metadata in beacons sent over Bluetooth can be decrypted if the TEK
is known, and detected if the correct metadata is not sent.
If such is detected, it can be suspected that a secret process is embedded in
the receiver's OS.

\item[\MARU{D} Reporting:]
Reporting must be digitally signed by a medical institute, and false reports
cannot be uploaded without collusion.
Because of blind signature, it is not possible to identify the person who has
tested positive (although identities can be inferred by the institute if only
a small number of people have tested positive there during a given period).

\item[\MARU{E} Stored reports:]
All reports will be proven for their existence, and it is not possible to
insert reports retrospectively.
If the application knows the TEKs, the user will be able to detect if
they are compromised by the OS and a false report without the consent of the
user is generated by a complicit medical institute.

\item[\MARU{F} Matching proximity IDs:]
Suppose that a second device that is carried with the phone constantly
collects beacons that would have been received by the phone.
If the rolling proximity ID generated from the reported
$\langle$TEK, base time$\rangle$ pair is compared to the IDs in the beacons,
and the user is notified of the exposure even though the IDs do not match,
then we can detect that a fraud is occurring in the OS.
However, this can also be detected without modifying GAEN.

\end{description}

\subsubsection{Newly introduced threats?}
Our proposal introduced two new parties:
medical institutes and application developers\footnote{Plus the public
database that tells certified public keys of medical institutes, but it should
be easily monitored by consumer groups and others.}.
We have to assume that these new third parties are also unreliable.

If the medical institute changes the key pair for each blind signature, it
can strip the reporter's anonymity, but the institute cannot do it alone
because the public keys have to be proven on a public database.
By colluding with the government, threats \ref{pro-collect-reporters} and
\ref{pro-agents} are made possible, and by colluding with OS developers or
phone vendors, threat \ref{pro-disable} is also made possible.
However, the increased number of parties that must collude makes it more
difficult to cheat than the original GAEN.

We propose to make the source code open to improve verifiability by consumer
groups and others, and to help protect privacy by allowing careful users to
build applications from the source code.
On the other hand, this makes it easier for malicious application developers
to introduce malware.
Nevertheless, the increased risk is related to phishing and whether users
install the wrong applications.
The risk of malicious developers themselves running rogue applications is the
same with the original GAEN.

\subsection{Potential Performance Impact}
We also roughly evaluate the extent to which the two designs can respond to an
increase in the number of tested-positive reporters.

Processes such as getting blind signatures from a medical institute before
reporting a positive test result, or calculating the Merkle tree from a set of
downloaded reports and querying the Merkle root stored in a public ledger,
are our additions to the original GAEN, but do not affect the overall
performance of the system as they are processed in a distributed manner,
although they do increase the power consumption of individual phones slightly.
We hope that the users will consider that this is the cost of better privacy.

On the server side, processing of uploaded $n$ reports involves the
following additional calculations in our proposal:
1) verification of the signature attached to the report (cost $O(n)$),
2) creation of the Merkle tree (cost $O(n)$; this takes about $2n$ digest
calculations),
and 3) storage of the Merkle root in the blockchain (cost $O(1)$).
The calculations can be parallelised and these costs can be load balanced by
adding processors (Merkle tree creation requires $O(\log n)$ sequential
processing).

\section{Related Work}\label{sec-related}
\subsection{Concerns on Exposure Notification}
The general ethical concerns and guidance on exposure notification and
contact tracing are well summarised in \cite{Ranisch2020:CTEthics}.

For security, \cite{cryptoeprint:2020:493} covers the types of attacks that
are possible.
\cite{hoepman2021critique} and \cite{Leith2020:GAEN} warn potential political
use of the tool.

The effectiveness of GAEN has been measured through experiments by
\cite{Wilson2020.07.17.20156539} and \cite{Leith_2020}.
Meanwhile, it was announced in February 2021 that the GAEN-based NHS COVID-19
application in UK has alerted 1.7 million contacts, and the UK government
estimates approximately 600,000 cases have been prevented since September
2020 \cite{GOV.UK2021:NHSApp}.
If this estimate is correct, GAEN is working effectively.

\subsection{Enhancements to GAEN Protocol}
A proposal to add GPS information to GAEN has been made by
\cite{raskar2020adding}.
While this may improve the accuracy of contact tracing, it raises concerns
about privacy.

\subsection{Blockchain-based Contact Tracing}
Some ideas on using blockchain for contact tracing have been proposed.
Many have proposed defining and running their own blockchain, but a ledger
system that starts small has a relatively large margin for malicious
involvement in replicating the state machine, making it difficult to provide
provability.
Many are also trying to deal with geographic information instead of or in
addition to Bluetooth proximity, which has potential difficulties in terms of
privacy protection, while proximity alone is proving to be effective enough
in reality.

Among such proposals, \cite{Xu2020:BeepTrace} and \cite{lv2020decentralized}
incorporate geolocation information.
\cite{Choudhury2020:CovidChain} proposes a potential intervention to privacy
with regard to promotion of public health.
\cite{Song2020:Blockchain} proposes to track users' travel trajectories.

\subsection{Safe Blues}
Safe Blues\cite{Dandekar2020:SafeBlues} simulates and predicts actual
infectious disease outbreaks by monitoring the spread of virtual viruses using
technology similar to device-based contact tracing.
As with actual exposure notifications, this seems feasible with privacy
protection using only proximity, provided that locality is taken loosely.
If this is the case, public health authorities may be able to proactively
combat infectious diseases by incorporating Safe Blues functionality into
exposure notification systems.
Even then, it is important to ensure that the system is verifiable by the
public, as we have shown in our proposal, so that it cannot be used by the
authorities to unfairly control the activities of the population.

\section{Conclusions}\label{sec-conclusions}

In this work, we have shown that minimal changes to GAEN, a working exposure
notification mechanism, can protect users' privacy without trusting third
parties.

Under our proposal, many of the privacy violations by OS developers,
smartphone vendors, medical institutes and government authorities could be
detected through verification by consumer groups and others.
This is not the case if they collude, but our proposal makes this more
difficult by increasing the number of actors that need to collude.

We believe that this will allow for safer contact tracing, and contribute to
healthier lifestyles for citizens who may want to or have to go out under
pandemic situations.

\bibliographystyle{plain}
\bibliography{exposure-notif}

\end{document}